# The InfraRed Imaging Spectrograph (IRIS) for TMT: Overview of innovative science programs

Shelley A. Wright[*a,b], James E. Larkin[c], Anna M. Moore[d], Tuan Do[a], Luc Simard[e], Maté Adamkovics[f], Lee Armus[g], Aaron J. Barth[h], Elizabeth Barton[h], Hope Boyce[a,i], Jeffrey Cooke[j], Patrick Cote[e], Timothy Davidge[e], Brent Ellerbroek[k], Andrea Ghez[c], Michael C. Liu[l], Jessica R. Lu[l], Bruce A. Macintosh[m], Shude Mao[n], Christian Marois[e], Mathias Schoeck[k], Ryuji Suzuki[o], Jonathan C. Tan[p], Tommaso Treu[q], Lianqi Wang[k], Jason Weiss[c], and IRIS team[k]

[a]Dunlap Institute for Astronomy & Astrophysics, Univ. of Toronto, Toronto ON, M5S 3H4 Canada;
[b]Dept. of Astronomy & Astrophysics, Univ. of Toronto, Toronto, ON, M5S 3H4 Canada;
[c]Physics & Astronomy Department, University of California Los Angeles, CA 90095 USA;
[d]Caltech Optical Observatories,1200 E California Blvd., Pasadena, CA 91125 USA;
[e]National Research Council of Canada - Herzberg, Victoria, BC, V9E 2E7 Canada;
[f]Astronomy Department, University of California Berkeley, CA 94720 USA;
[g]Infrared Processing and Analysis Center, California Institute for Technology, Pasadena, CA 91125;
[h] Department of Physics & Astronomy, University of California Irvine, Irvine, CA 92697 USA;
[i]Department of Physics & Engineering, University of Saskatchewan, SK S7N 5A1 Canada;
[j]Centre for Astrophysics, Swinburne University of Technology, Hawthorn, VIC 3122, Australia;
[k]Thirty Meter Telescope Observatory Corporation, Pasadena, CA 91105 USA;
[l]Institute for Astronomy, University of Hawaii, Manoa, HI 96822 USA;
[m]Department of Physics, Standford University, Stanford, CA 94305 USA;
[n]National Astronomical Observatories, Chinese Academy of Sciences, Beijing 100012, China;
[o]National Astronomical Observatory of Japan, Osawa, Mitaka, Tokyo, 181-8588 Japan;
[p]Departments of Astronomy and Physics, University of Florida, Gainesville, Florida 32611, USA;
[q]Department of Physics, University of California, Santa Barbara, CA 93106-9530, USA

## ABSTRACT

IRIS (InfraRed Imaging Spectrograph) is a first light near-infrared diffraction limited imager and integral field spectrograph being designed for the future Thirty Meter Telescope (TMT). IRIS is optimized to perform astronomical studies across a significant fraction of cosmic time, from our Solar System to distant newly formed galaxies (Barton et al. [1]). We present a selection of the innovative science cases that are unique to IRIS in the era of upcoming space and ground-based telescopes. We focus on integral field spectroscopy of directly imaged exoplanet atmospheres, probing fundamental physics in the Galactic Center, measuring $10^4$ to $10^{10}$ $M_\odot$ supermassive black hole masses, resolved spectroscopy of young star-forming galaxies ($1 < z < 5$) and first light galaxies ($6 < z < 12$), and resolved spectroscopy of strong gravitational lensed sources to measure dark matter substructure. For each of these science cases we use the IRIS simulator (Wright et al. [2], Do et al. [3]) to explore IRIS capabilities. To highlight the unique IRIS capabilities, we also update the point and resolved source sensitivities for the integral field spectrograph (IFS) in all five broadband filters (*Z, Y, J, H, K*) for the finest spatial scale of 0.004″ per spaxel. We briefly discuss future development plans for the data reduction pipeline and quicklook software for the IRIS instrument suite.

---

[*] Send correspondence to wright @ astro.utoronto.ca

# 1. INTRODUCTION

The coming decade promises a revolution in astronomical discoveries from new instruments on upcoming telescopes like the Giant Segmented Mirror Telescopes (GSMTs), James Webb Space Telescope (JWST[4]), and Large Synoptic Survey Telescope (LSST[5]). IRIS (InfraRed Imaging Spectrograph) is one such revolutionary instrument being designed to sample the diffraction limit obtained from the multi-conjugate adaptive optics system, NFIRAOS[6], on the future Thirty Meter Telescope (TMT[7]). IRIS will house a dedicated near-infrared (0.845 – 2.4 μm) imager and integral field spectrograph (IFS), as described by Moore et al. [8], this conference. Both imager and IFS will offer unprecedented angular resolution, sampled with the imager at 4 mas/pixel with a total field of view[†] of 16.4″ × 16.4″ and the IFS with four spatial scales spanning 4 – 50 mas/spaxel with a range of field of views from 0.064″ × 0.51″ to 4.4″ × 2.25″. The IFS spectral resolution has been carefully selected to maximize scientific return at R=4,000 and include an additional set of high spectral resolution gratings at R=8,000.

IRIS is being optimized to perform astronomical studies of point and resolved sources with a wide range of surface brightness. We have investigated a range of science cases and used the IRIS data simulator[2,3] to facilitate the conceptual design of the instrument. These cases include Solar System objects, exoplanets, microlensing, star-forming regions, the Galactic Center, nearby galaxies and supermassive black holes, strong gravitational lensing, high redshift galaxies and quasars, and first-light galaxies (see Barton et al. [1]).

We present a selection of innovative IRIS science cases that are uniquely suited to studies with TMT and NFIRAOS and the expected powerful synergy with other ground and space-based telescopes in the coming decade. All three GSMTs with diffraction limited capabilities will be able to exploit a range of interesting science cases, and each of these GSMTs plan on a first-light diffraction limited IFS (GMTIFS[9] for GMT and HARMONI[10] for E-ELT). Compared to future space-based missions, JWST will have an IFS with similar wavelength coverage (NIRSPEC[11]), but will have limited spatial resolution at 0.1″/spaxel and spectral resolution of R < 2700 and limited lifetime. IRIS imager and spectrograph are being designed for high photometric and astrometric accuracy, which will open a new paradigm in near-infrared experimental astrophysics. In this paper, we present the following sample of science cases that are particularly exclusive to the capabilities to IRIS and NFIRAOS: directly-imaged exoplanet atmospheres, the Galactic Center, supermassive black holes, high-redshift star-forming galaxies, first light galaxies, and strong gravitational lensed systems.

# 2. DIFFRACTION LIMITED IFS: POINT AND RESOLVED SENSITIVITIES

Our team has developed an end-to-end data simulator for the imager and spectrograph to assess the capabilities of IRIS and to aid in the development of the data reduction pipeline, see Wright et al. [2], Do et al. [3]. We have continued to update the simulator as the design of IRIS and NFIRAOS have evolved, modifying the point spread functions (PSFs) from NFIRAOS and total throughput per waveband across the entire system[8]. Figure 1 presents the sensitivity for both resolved and point sources on the IFS for five of the IRIS broadband filters, $Z$ ($\lambda_{cen}$=0.93 μm), $Y$ ($\lambda_{cen}$=1.09 μm), $J$ ($\lambda_{cen}$=1.27 μm), $H$ ($\lambda_{cen}$=1.63 μm), and $K$ ($\lambda_{cen}$=2.18 μm), in its finest spatial scale of 0.004″ per spaxel. Our simulations show that the IFS using 0.004″/spaxel to observe a 25 mag ($Z, Y, J, H,$ or $K;$ Vega) point source in 5 hours (20 × 900s) will be able to achieve a signal-to-noise ratio (SNR) of ~10 per wavelength channel (R=4000) for a 2λ/D aperture (given a flat spectrum). In addition, the IFS using 0.004″/spaxel on a resolved source with a uniform surface brightness in 5 hours (20 × 900s) will be able to reach a SNR of ~10 per wavelength channel per spaxel for a $Z$=15.5 mag arcsec$^{-2}$ and $K$=14 mag arcsec$^{-2}$. Thus the sensitivity estimates for NFIRAOS and IRIS in Figure 1 are unprecedented compared to current facilities.

---

[†] The IRIS imager is currently designed to have a 16.4″ × 16.4″ field of view, but the technical team has developed a plan for an easy upgrade to a larger field of view of 32.8″ × 32.8″.

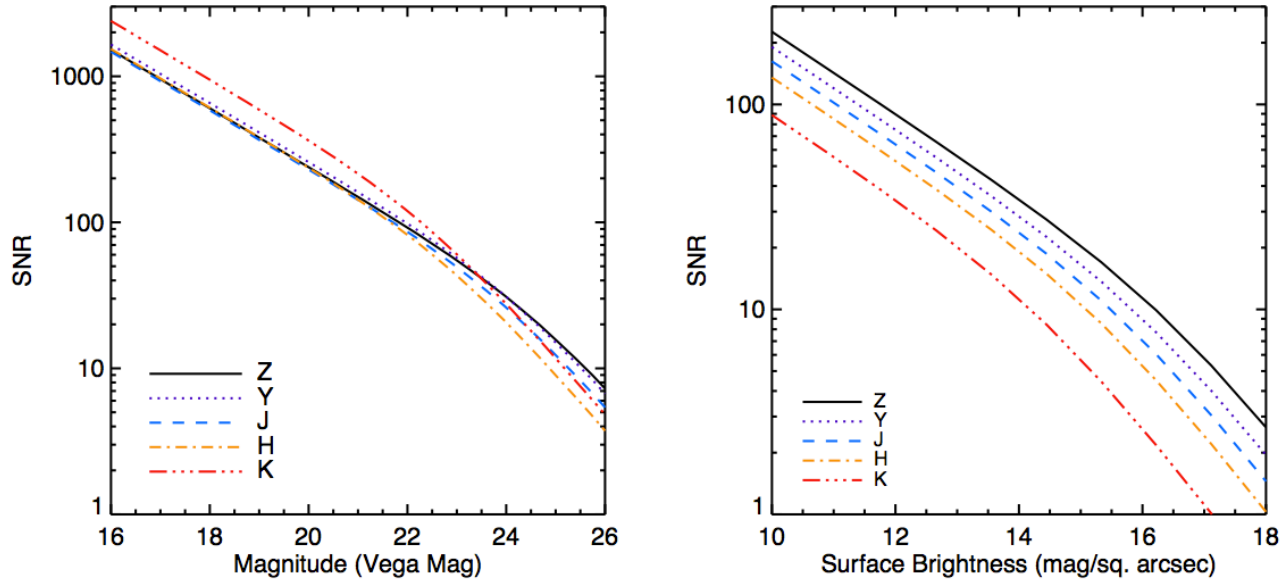

**Figure 1**: IRIS IFS point source (left) and surface brightness (right) signal-to-noise ratios (SNRs) per spectral channel at R=4000 for a 5 hour integration (20 frames × 900 second exposures) using the finest spatial scale of 0.004″ per spaxel. All magnitudes are in Vega units. All five broadband filters are shown: $Z$ ($\lambda_{cen}$=0.93 μm), $Y$ ($\lambda_{cen}$=1.09 μm), $J$ ($\lambda_{cen}$=1.27 μm), $H$ ($\lambda_{cen}$=1.63 μm), and $K$ ($\lambda_{cen}$=2.18 μm). The sensitivities are calculated using the IRIS simulator and latest NFIRAOS PSFs. Compared to resolve sources, point source sensitivities change with wavelength due to improved adaptive optics correction at longer wavelengths.

## 3. SELECTION OF IRIS SCIENCE CASES

### 3.1 Moderate resolution spectroscopy of directly imaged exoplanets

With over 1000 planetary systems now identified, the ubiquity of planets in our Galaxy has been spectacularly confirmed.[12,13] Much of the focus of the next decades will now shift to characterization of planetary systems in order to understand the way in which they form and evolve. Spectroscopy of exoplanets offers a powerful tool to not only constrain their atmospheric properties as a function of mass, age, and separation, but also to reveal the fingerprints of planet formation. With high enough spectral resolution, individual atomic and molecular features can be resolved to perform detailed abundance analysis that can offer clues about the composition of exoplanets.[14,15] IRIS astrometric accuracy in the imager mode will also play a crucial role in astrometric monitoring of directly imaged planets and 3D orbital analysis, that will be coupled with the radial velocity measurements from spectroscopy.

Direct imaging is the latest technique for detecting and studying extrasolar planets (e.g., HR 8799[16], Beta Pic[17], HD95096b[18]). This technique has the distinct advantage, compared to the Doppler radial velocity and transit detection, of being able to directly measure the photometric and spectroscopic properties of the planets themselves. With the recently commissioned Gemini Planet Imager (GPI[19]) and SPHERE[20] on VLT a new population of widely separated (~5-100 AU) extrasolar planets are expected to be discovered in the coming years. However, characterization will initially be limited to low spectral resolution (R~50-100) by the instrumental design (GPI and SPHERE), scattered light artifacts (current-generation AO), and sheer faintness. At the resolutions of 4000 or more offered by IRIS, multiple absorption lines from neutral atoms (e.g. Na and K) and molecules (e.g. FeH and VO) can be resolved and used to assess surface gravity (a proxy for age and mass) and metallicity (a key diagnostic for the formation of gas-giant planets). In addition, detailed modeling of the dominant opacity sources of $H_2O$, $NH_3$, $CO$, and $CH_4$ will provide more accurate temperature measurements and also probe non-equilibrium chemistry though the relative abundance of CO to $CH_4$, providing important clues on planetary formation.[21,22,23] Figure 2 provides an example of the quality of spectroscopy that will be routinely achievable with IRIS on directly imaged exoplanets. The higher SNR and photometric accuracy achieved with IRIS spectroscopy on directly imaged planets, compared to GPI and SPHERE, will be important for studying cloud variations and atmospheric weather, as well as spin rotation.[24] IRIS will therefore offer an unequaled probe of the physics of planetary atmospheres compared to current and upcoming facilities.

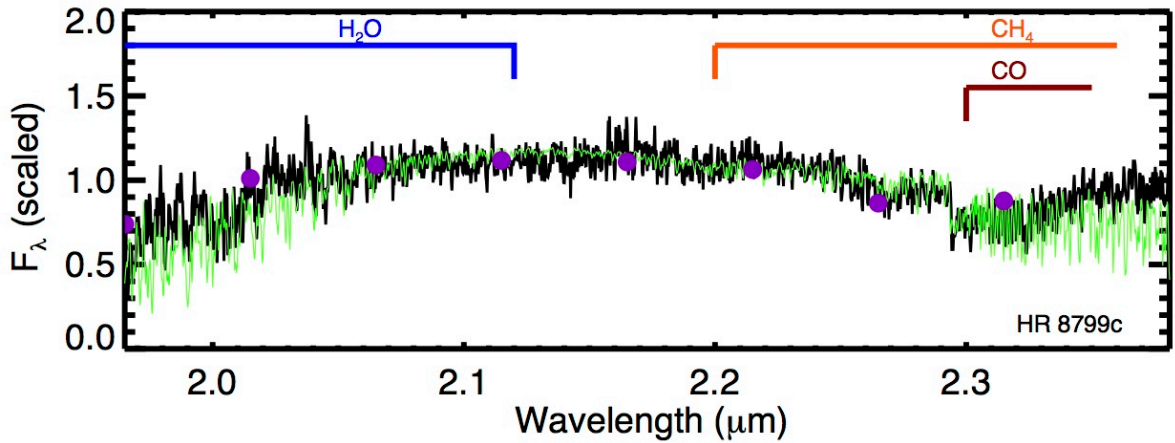

**Figure 2**: The observed R=3800 spectrum of the HR 8799c planet (black) from OSIRIS at Keck Observatory (Konopacky et al. [23]). The green curve represents the best-fit atmospheric model. A binned version of this spectrum (R~10, purple dots) illustrates what the spectrum would look like if it had the typical resolution of space-based transit spectroscopy. This is one of the highest spectral resolution observations of any extrasolar planet atmosphere. The R=3800 resolution allowed for detailed measurements of the strengths of the molecular features $H_2O$ and CO, and the lack of $CH_4$. This is one of the few targets observable with current instrumentation due to its brightness and relatively wide separation from the host star[16]. IRIS will be able to routinely achieve this level of spectroscopy on the newly discovered extrasolar planets at lower masses, cooler temperatures, and closer separations.

### 3.2 Crowded field spectroscopy: resolved stellar populations at the Galactic Center

One of the leading science cases for IRIS is to study the Galactic Center with unprecedented sensitivity and spatial resolution. The Galactic Center is the closest laboratory for studying the environments and fundamental physics of supermassive black holes and offers several unique science cases for TMT, e.g., Yelda et al. [25]. Both the IRIS imager and IFS have been carefully designed to characterize the surroundings of the supermassive black hole (SMBH), SgrA*, at the Milky Way's center.[26,27] The high relative astrometric accuracy of 30 μas will offer a test of General Relativity and probe the distribution of dark matter through orbital monitoring of the stars surrounding SgrA*.[28,29] Characterizing the stellar population at the Galactic Center is of great interest to ascertain how stars form and evolve around a SMBH.[30,31] The young stellar population near SgrA* has puzzled astronomers, as young massive stars should have difficulty forming in close proximity to a SMBH. Researchers have thus far been limited to studying only the most luminous stars in the area, including OB main sequences stars, red giants, and Wolf-Rayet stars. Currently, OSIRIS[32] on Keck is able to achieve spectroscopy with sufficient SNR to measure spectral types and radial velocities for Kp < 15.5 mag stars. In contrast, IRIS is predicted to have the sensitivity to allow for high SNR spectroscopy on Kp = 20 - 21 mag stars (Figure 3). These sensitivities and high angular resolutions will allow researchers to study the low mass-end of the main sequence, which will be crucial for investigating the stellar population. They will also provide essential radial velocity measurements to couple with the proper motion monitoring via imaging for deriving 3D orbital solutions. Currently, the shortest period star identified around the black hole has a period of 11.5 years.[33] IRIS is expected to identify multiple sources with shorter orbital periods (1-2 years), which will be instrumental for fundamental physics studies like testing General Relativity.

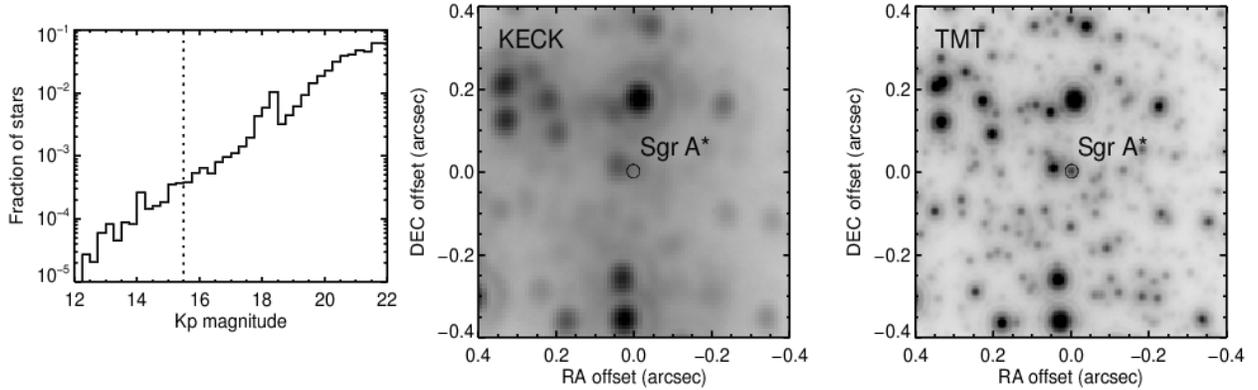

Figure 3: We present the theoretical luminosity function of the stellar population within 1″ of Sgr A* including both the young and old stellar population in the Galactic Center that has been explored by Lu et al. [31] and Do et al. [34], respectively. The fraction of all stars as a function of Kp magnitude (left) illustrates the number of stars observable with TMT (right) compared to the current capabilities of Keck AO imaging (middle). The Keck LGS-AO has a magnitude limit of 15.5 magnitude for spectroscopy (dotted line in left figure), while TMT will easily observe low mass stars on the main sequence and probe the initial mass function, which is essential for understanding the mysterious origins of the stellar population surrounding the SMBH. The IFS will be able to do spectroscopy with a signal-to-noise ratio of 20 in 2 hours of observations for Kp=20 mag stars. This is important for investigating the star formation history, as well as coupling radial velocity measurements for 3D orbital analysis to investigate General Relativity and the dark matter distribution.

### 3.3 Black hole mass measurements: probing $M_{BH}$-σ and $M_{BH}$-L

Understanding the interplay between supermassive black hole (SMBH) growth and formation of its host galaxy is currently one of the most outstanding astrophysical questions. Some of the largest puzzles are the origins of the $M_{BH}$-σ relationship,[35,36] the role of active galactic nuclei (AGN) feedback, mergers of black holes and their effect on the centers of galaxies, and how to effectively fuel black hole growth. The angular resolution of IRIS represents the next major leap in observational capabilities for SMBH detection. These measurements will greatly impact our understanding of galaxy and SMBH formation. TMT will make it possible to expand the mass range over which black holes can be detected, into the largely unexplored low to high-mass range of the $M_{BH}$-σ relation (see Figure 4). Quiescent galaxy black hole masses have been primarily measured from stellar or gas dynamics, with only ~80 galaxies measured to-date[36]. Most recently, the use of adaptive optics on 8-10m class telescopes has allowed some of the most precise measurements of SMBH masses,[38,39,40] since the angular resolution is sufficient for sampling the sphere of influence of the SMBH and near-infrared wavelengths is less susceptible to dust extinction. Do et al. [3] have recently explored the prospects for measuring SMBH masses in the range of $10^4$ to $10^{10}$ $M_\odot$ using IRIS. They demonstrated that IRIS will provide the SNR to easily measure intermediate SMBH masses (>$10^6$ $M_\odot$) out to z~0.2 and will have the capability of investigating the elusive intermediate black holes in globular clusters and dwarf galaxies. In contrast, the spatial scales offered by JWST and HST are insufficient to probe the sphere of influence of the SMBH, which is necessary to make mass measurements and explore the full $M_{BH}$-σ phase space. It is also worth noting that the most massive black holes ($10^{9-10}$ $M_\odot$) have the largest sphere of influence, and IRIS has the angular resolution necessary to resolve the stellar or gas dynamics around such behemoth black holes even early in the universe (z > 1). These would be incredibly challenging observations, but TMT's angular resolution makes this project potentially viable.

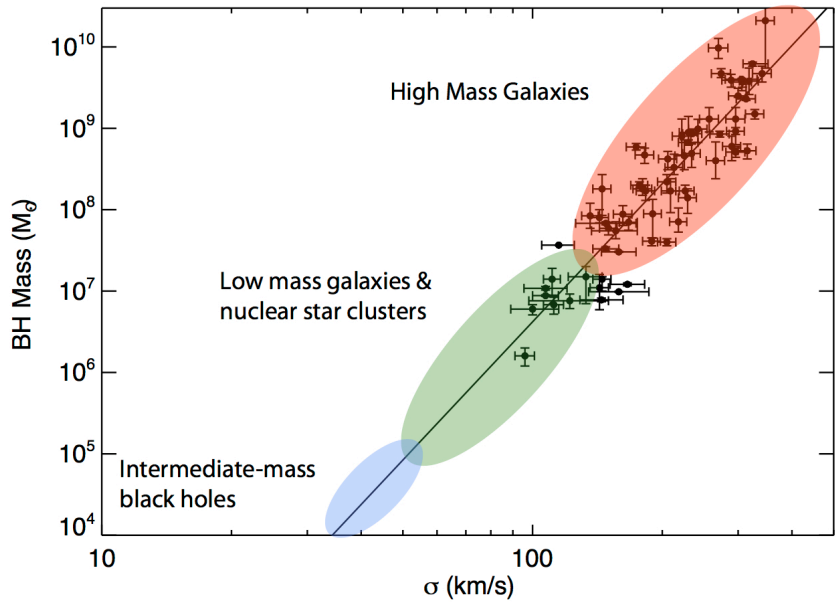

**Figure 4:** The current black hole mass and stellar velocity dispersion ($M_{BH}$ vs. $\sigma$) phase space (values obtained from McConnell & Ma [37]). The highlighted regions show the capabilities for measuring black hole masses with TMT and IRIS. For black hole masses in the range $10^6$ - $10^7$ $M_\odot$, very few detections exist outside of the Local Group. TMT will make it possible to resolve the black hole sphere of influence in nearby late-type spiral galaxies and dwarf ellipticals (red and green regions), allowing measurements that were previously only possible for Local Group galaxies to reach greater distances (see Do et al. [3]). IRIS will even be able to explore the elusive intermediate black hole mass ($10^3$ to $10^4$ $M_\odot$) phase space (shaded blue).

### 3.4 Resolved spectroscopy of high-redshift galaxies (1 < z < 5)

IRIS will expand our knowledge of galaxy formation in the young universe. While we are beginning to compile measures of the global parameters (e.g., luminosity, color, star formation rate, gas and dust content, and stellar mass) of very high redshift galaxies, there is still a gap in our knowledge of the processes that regulate galaxy growth and evolution even at modest redshifts of z ~1 – 3. This is an extremely important epoch in the lives of normal galaxies, mirrored in the precipitous drop in the cosmic star formation rate density below z ~ 2.[41] Recent groundbreaking observations have made use of IFSs and AO (e.g., OSIRIS and SINFONI) on 8-10m class telescopes, and have been able to probe the dynamical processes of individual high-redshift galaxies (z > 1.5),[42,43,44,45] but they are limited to the most luminous galaxies (10 – 100 $M_\odot$ yr$^{-1}$) and coarse sampling ($\geq$ 0.5 kpc) compared to the sampling capabilities of IRIS and TMT (e.g., 34 pc at z = 2).

IRIS will be able to spatially resolve nebular emission lines such as H$\beta$, [O II], [O III], [N II], and [S II] in addition to H$\alpha$. With these measurements, it will be possible to construct spatially resolved maps of diagnostic nebular line ratios on a range of galaxy masses and redshifts. In addition to spatially isolating regions dominated by different excitation mechanisms (photoionization, shock heating, etc.), the spatial resolution of IRIS at high-z will also probe the star formation history and differential enrichment of large H II regions and individual super star clusters. It will therefore be possible to compare the abundances of different components of high redshift galaxies, differentiating (for instance) between a central AGN,[46,47] bulge, and the nascent stellar disk. This will offer strong constraints on the origin of the mass-metallicity relation[48,49] and the primary sources of feedback responsible for regulating star formation during the epoch when the majority of stars in the visible universe were formed.

IRIS will also be able to probe lower integrated star formation rates, well below 1 $M_\odot$ yr$^{-1}$. IRIS will achieve sensitivities that are 20-30 times better than current 8-10m telescope AO-fed IFSs. This is illustrated in Figure 5, where the 0.05″ scale will allow observations of z = 1.5 star-forming galaxies with integrated star formation rates of 0.1 – 10 $M_\odot$ yr$^{-1}$. The 0.025″ and 0.05″ scales are advantageous for detecting fainter emission lines (e.g., [OII], H$\beta$, [OIII], [NII]) to study the 2D metallicity and dynamics of high-redshift galaxies (>10 $M_\odot$ yr$^{-1}$), while the finer spatial scales 0.004″ and 0.009″ will be used to study luminous star forming galaxies at high spatial resolution. Our team has already explored sensitivities and this science case for high-redshift star-forming galaxies in greater detail (see Wright et al. [2]).

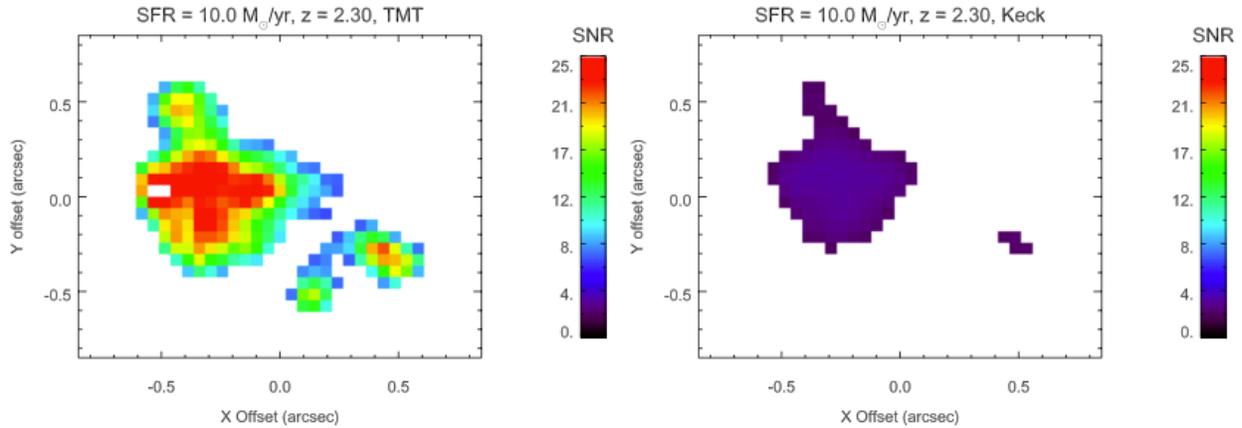

**Figure 5:** A simulated z = 2.3 galaxy, with Hα redshifted to K-band and an integrated star formation rate of 10 $M_\odot$ yr$^{-1}$ (left). This is compared to Keck observations of the same simulated galaxy (right) using the same spatial sampling of 0.05″ per spaxel. Both simulations were for a R=4000 IFS with a 2 hour (8x900s) integration in K-band. Complex structure with multiple knots can be seen with IRIS and TMT, while only the brightest portion of the galaxy is barely detectable (SNR~3) with Keck. The higher SNRs for IRIS are also essential for detecting other emission lines, like [NII], [OIII], [SII], for diagnostics of their ionization parameters and chemical abundances. The finest spatial scale of 0.004″ and 0.009″ will be able to probe down to individual giant molecular cloud regions in this galaxy.

Characterizing Lyα emitters (LAEs) and Lyman continuum at moderate redshifts (z = 2 - 4) will also be essential for understanding the epoch of reionization (z > 6). The Lyman continuum flux escaping from high-z galaxies is responsible for the bulk of the ionizing photons necessary to initiate and maintain reionization.[50,51,52] Theoretical interpretations that predict the fraction of ionizing photons are based on the observed relationship between the escape fraction of Lyman continuum and optical nebular emission lines, such as [OII], [OIII], and Balmer lines.[53] Galaxies with the highest observed fraction of escaping Lyman continuum flux are faint, blue galaxies and the LAEs that have typical star formation rates of < 1 $M_\odot$ yr$^{-1}$.[54,55] Therefore, the sensitivity of TMT+IRIS is necessary to detect faint high-z nebular emission lines to directly calibrate against Lyman continuum measurements. Empirically establishing these relationships at z = 2 - 4 will be essential for deducing the ionizing escape fraction in the peak era of reionization (see Section 3.5). The angular resolution of IRIS will enable an understanding of the physical mechanisms and conditions that allow the Lyman continuum to escape in high-z galaxies, and therefore its subsequent effect on galaxy formation.

### 3.5 Ly-α spectroscopy of first light galaxies (6 < z < 12)

Understanding the first galaxies in the Universe is a crucial problem in cosmology and astrophysics. The epoch of "first light" – between 300 to 900 million years after the Big Bang – is a key period in cosmic evolution for understanding the first seeds of stellar, supermassive black hole, and galaxy formation. It is the epoch where we can probe the building blocks of the first galaxies and understand their stellar population and intergalactic medium. The science case for "first light" galaxies is one of the prime drivers for upcoming facilities like JWST and the GSMTs.

JWST will have unprecedented sensitivity that will allow for the discovery of first light galaxies.[56] Through the use of accurate photometric redshifts, JWST will likely discover the most distant and luminous galaxies yet observed. IRIS will be uniquely positioned to conduct spatially resolved near-infrared spectroscopy of these distant young galaxies. Resolved spectroscopy will be essential for spatially resolving rest-frame ultraviolet emission lines, like Lyα, from the star-forming regions of the galaxies and/or any potential AGN. Observations of the Lyα line profile will reveal a great deal about the dynamics of the gas in these extremely young galaxies. In many redshift intervals, the He II (164.0 nm) line is also accessible. If a strong HeII feature is discovered, the observations will provide compelling evidence for a hard radiation field from massive stars formed from low metallicity gas, thus helping to map out the chemical enrichment histories of these elusive galaxies.

IFS observations of the Lyα emission line will be crucial for tracing the kinematics and morphologies of first light galaxies. The spatial distribution will potentially allow us to rule out AGN, and the line profile of the Lyα feature will yield strong constraints on the ionization state of the nearby intergalactic medium. Although NIRSPEC on JWST has an IFS, the spatial sampling (0.1″ per spaxel, R<2700) is not as well suited for the sizes of these first light galaxies compared to IRIS with TMT. The surface brightness profiles of the most distant galaxies observed to-date (z = 7 - 8) have typical half light radii of 0.4 – 0.8 kpc (0.09 – 0.18″ at z=9).[58] Therefore to resolve the internal structure of first light galaxies it is necessary to have high angular resolution and sensitivity to observe the velocity structure and widths of Lyα. Our team has explored a range of expected Sersic indices (n = 1 - 4) and integrated light fluxes of Lyα at varying redshifts.[58] Figure 6 illustrates how the SNR is impacted by the surface brightness profiles (total flux of $0.1–2\times10^{-18}$ erg s$^{-1}$ cm$^{-2}$) of Lyα across a z=8.9 first light galaxy. The "clumpier" the surface brightness profiles the higher SNR achieved on these sources, as has been observed in present IFS observations of high-z galaxies.

The detectability of Lyα emission will largely be dominated by the escape fraction and optical depth to Lyα. Recent spectroscopic studies of z = 8 Y-dropout candidates imply that the optical depth at these redshifts may already be high enough to limit their detection.[59,60] Treu et al. [59] used MOSFIRE[61] to reach a 5σ limit of $0.4-0.6\times10^{-17}$ erg s$^{-1}$ cm$^{-2}$, which is still 2 to 20 times brighter than sources detectable with IRIS. However, even if the optical depth is considerably higher at z > 8, observing redder lines (e.g., CIII 190.9 nm) may prove to be a better probe of first light galaxies. For instance, CIII emission emanating from z = 8 - 10 galaxies is redshifted into the J, H, and K bands, where the NFIRAOS performance is best. Although, CIII is intrinsically fainter than Lyα, it may still have an increased SNR based on lower extinction and better instrumental performance of IRIS and NFIRAOS at these longer wavelengths. IRIS and TMT will provide the necessary sensitivity and spatial resolution to sample the dynamics of first light galaxies. This is a key science driver for all the GSMTs, and IRIS with NFIRAOS is well suited for this science objective.

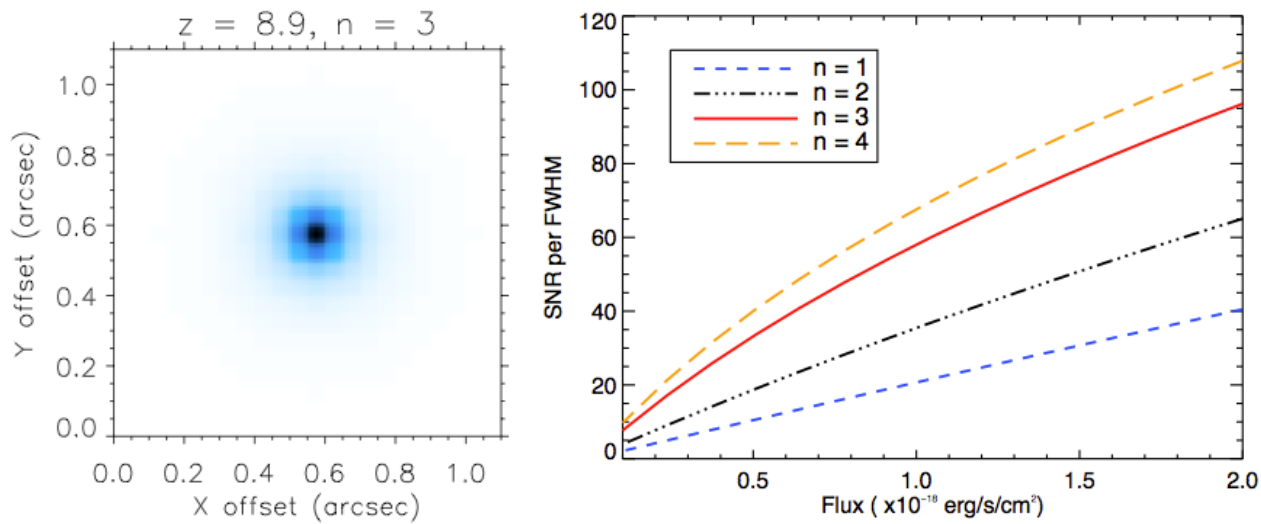

**Figure 6**: We investigated IFS sensitivities of Lyα emission from young distant galaxies with varying surface brightness profiles and redshifts (6 < z < 12). Using the measured half light radius of first light galaxies, between 0.4 – 0.8 kpc (Oesch et al. [56]), we explored a range of Sersic profiles and the SNR per FWHM versus the total Lyα flux of the galaxy[51]. (LEFT) We present a simulated image of a z=8.9 galaxy with Lyα of n=3 Sersic index with normalized flux with a half-light radius of 0.6 kpc. (RIGHT) A range of Sersic profiles with their corresponding SNR per FWHM is presented for the same galaxy. This is using the 0.05″ IFS plate scale at R=4000 with a Lyα velocity dispersion of 25 km s$^{-1}$ observed for a total of 5 hours (20 × 900s) in H-band.

### 3.6 Strong gravitational lensed systems: investigating first light galaxies and measuring dark matter substructure

Gravitational lensing will play an essential role in studying the most distant galaxies and will be crucial for mass estimates of galaxies at intermediate redshifts. The combination of an IFS and AO on GSMTs is ideally suited for resolved spectroscopy of lensed z = 8 - 10 galaxies. First light gravitational lensed sources have yet to be discovered, but hundreds of high-redshift (z > 1) strong lenses will be found by future surveys with JWST and LSST. A recent example is the discovery of a lens at z = 1.5 using HST.[62] Resolved spectroscopy using IRIS and TMT will provide the necessary angular resolution to decouple the emission of the lens and background galaxy, even in the most complicated systems.

These observations will provide measurements for both the lens (z = 1 - 2 massive galaxy) and lensed source (z = 8 - 10 galaxies of Lyα emission) with a direct estimate of the total mass of the lens and high angular resolution velocity profiles of Lyα of the lensed galaxy. For instance, for a z=9 un-lensed galaxy, the angular size scale of 0.05″ corresponds to 228 pc. Assuming a typical lensing magnification factor of 10, the spatial sampling in the source plane will be 22.8 pc per spaxel. For the 0.009" IFS plate scale, the sampling for lensed z = 9 galaxy would be 4.1 pc, which remarkably, is equivalent to ground-based observations of Virgo cluster galaxies.

Strong gravitational lensed systems also offer powerful probes of the substructure in dark matter halos at a range of redshifts. With the unprecedented resolution of IRIS, we will be able to probe dark matter substructures in lensing galaxies with mass a factor ~100 lower than current measurements, which will provide a stringent test of a generic prediction of the Λ Cold Dark Matter model. Local studies of nearby galaxies and their sub-halos rely heavily on baryonic physics (e.g. tidal streams[63]), whereas strong gravitational lensing is able to directly probe the dark matter distribution and measure lower mass sub-halos. Photometric and astrometric studies have been used to investigate the gravitational potential of the lens system based on the source image time delay, position, and anomalous flux ratios.[64,65,66,67] Recently, Keck OSIRIS AO observations have been able to use observed flux differentials in a quadruple lensed background QSO (z=3.6) to measure the mass of sub-halos in a z=0.4 lens elliptical galaxy (see Figure 7).[68] As previously stated, thousands of quadruple lensed systems will be discovered with JWST and LSST at even greater redshifts (z = 0.5 – 2). With the use of TMT and IRIS we will therefore be able to extend current studies to a larger number of systems at higher redshifts and with lower mass sub-halos (~$10^6$ $M_\odot$).

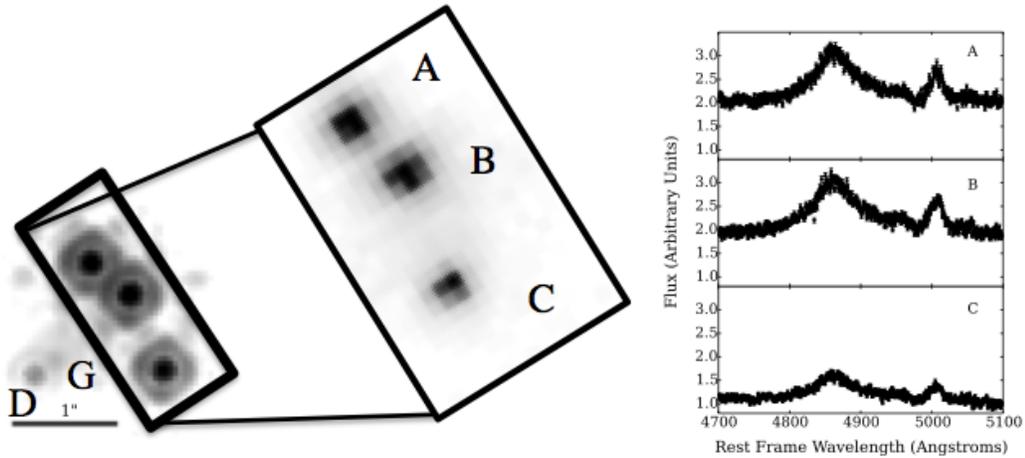

**Figure 7:** Results of narrow-line emission analysis using Keck AO with OSIRIS on a gravitational lensed z=3.6 QSO (Courtesy Nierenberg et al. [68]). (LEFT) HST NICMOS F160W image of B1422+231. (MIDDLE) OSIRIS field of view of three lensed images (A, B, C) in [OIII] emission, which were used to measure differential flux ratios between the lensed images to derive the mass of sub-halos in the lens system (z=0.4 elliptical galaxy). (RIGHT) OSIRIS extracted spectra for each lensed image. This was the first demonstration of using an AO + IFS to measure sub-halo masses with flux anomaly measurements in quadruply lensed system.

## 4. DATA REDUCTION AND QUICKLOOK

IRIS will have a dedicated data reduction pipeline and quicklook visualization tools for both the IFS and imager. These data reduction pipelines will be designed to run both in real-time for observational acquisition and data quality assessment during an observing sequence, while also offering more advanced post-processing analysis and reductions. Our team has begun to look at the data reduction procedures needed for the IRIS instrument suite. A data reduction pipeline refers to processing all "raw" files (2D frames with data numbers per second) generated during an observational or calibration sequence. The data reduction pipeline for the IFS will need to address the unique spectral extraction techniques for both the lenslet and slicer based spectrograph.[8]

The imager data reduction pipeline will include typical reduction procedures of sky and dark subtraction, correction of detector artifacts (e.g., crosstalk), correction of cosmic rays, flat fielding, field distortion correction, flux calibration,

PSF calibration, and advanced shift and add mosaicking routines. The expected challenges for this pipeline will likely be addressing new artifacts and anomalies with the newest Hawaii-4RG detectors and PSF calibrations.

There are several pipelines for existing near-infrared IFSs (like OSIRIS[31], SINFONI[69], GPI[70]) on which to base the IRIS reduction procedures. The IFS data reduction pipeline will use similar procedures as those described for the imager, with additional routines for flat fielding, spectral extraction, wavelength solution, assembly of the data cube ($x$, $y$, lambda), and atmospheric dispersion correction for any residuals.

Both day and night time calibration data will need to be obtained. During the daytime, dark frames, arc lamp for wavelength solution and resolution, white light illumination for flats and IFS spectral extraction, white light fiber for image quality, and a pinhole grid for optical distortion solutions will be obtained. Other "metadata" needed for the data reduction pipeline will be well organized and easily accessible for all day-time calibration frames, NFIRAOS configuration, PSF calibration and telemetry information (e.g., real-time seeing monitor information), and all environmental information for the telescope and atmosphere. Our team will make use of existing near-infrared pipelines for resources, but we will also implement new algorithms based on the lessons learned from these pipelines. Ease of use and fast-processing at the telescope is essential for the success of IRIS.

## 5. SUMMARY

IRIS is being designed to provide a broad spectrum of scientific capabilities that will exploit the first-light capabilities of TMT to answer some of the most fundamental questions in astrophysics. We have presented a small sample of science programs that are uniquely suited to IRIS, maximizing the angular resolution and sensitivity offered by TMT. In particular, we have highlighted IRIS science cases for (1) studying exoplanet atmospheres at moderate spectral resolution and their orbits from astrometric monitoring; (2) spectroscopy to study the stellar population and test General Relativity at the Galactic Center; (3) building up crucial statistics for measuring supermassive black hole masses; (4) resolved spectroscopy of high-redshift galaxies and (5) first light galaxies in the epoch of reionization; and (6) measuring dark matter substructure with the use strong gravitational lens systems. For each of these cases, we have utilized the IRIS data simulator to present realistic estimates of the performance of the instrument. These simulations show that IRIS will enable unique and unprecedented observations that will shape our understanding of both the nearby and distant Universe.

## ACKNOWLEDGEMENTS

The TMT Project gratefully acknowledges the support of the TMT collaborating institutions. They are the Association of Canadian Universities for Research in Astronomy (ACURA), the California Institute of Technology, the University of California, the National Astronomical Observatory of Japan, the National Astronomical Observatories of China and their consortium partners, and the Department of Science and Technology of India and their supported institutes. This work was supported as well by the Gordon and Betty Moore Foundation, the Canada Foundation for Innovation, the Ontario Ministry of Research and Innovation, the National Research Council of Canada, the Natural Sciences and Engineering Research Council of Canada, the British Columbia Knowledge Development Fund, the Association of Universities for Research in Astronomy (AURA), the U.S. National Science Foundation and the National Institutes of Natural Sciences of Japan.